\def\g5{\gamma_5}

\documentclass{elsart}

\usepackage{epsfig}

\begin{document}

\begin{frontmatter}

\title{New Measurement of Parity Violation in Elastic Electron-Proton 
Scattering and Implications for Strange Form Factors}

\author[calstate]{K.~A.~Aniol},
\author[wandm]{D.~S.~Armstrong},
\author[wandm]{T.~Averett},
\author[saclay]{M.~Baylac},
\author[saclay]{E.~Burtin},
\author[newhamp]{J.~Calarco},
\author[princeton]{G.~D.~Cates},
\author[saclay]{C.~Cavata},
\author[mitlns]{Z.~Chai},
\author[maryland]{C.~C.~Chang},
% ...

%\address{}

%\author{\vspace*{\bigskipamount}\centerline{(HAPPEX Collaboration)}} 

\address[calstate]{California State University - Los Angeles,Los Angeles, California 90032, USA } 

% ...
\address[maryland]{University of Maryland, College Park, Maryland 20742, USA} 

% ...
\address[mitlns]{Massachusetts Institute of Technology, Cambridge, Massachusetts 02139, USA} 

\address[newhamp]{University of New Hampshire, Durham, New Hampshire 03824, USA} 

% ...
\address[princeton]{Princeton University, Princeton, New Jersey 08544, USA} 

% ...
\address[saclay]{CEA Saclay, DAPNIA/SPhN, F-91191 Gif-sur-Yvette, France } 

% ...
\address[wandm]{College of William and Mary, Williamsburg, Virginia 23187, USA}

\date{\today}

\begin{abstract} 

We have measured the parity-violating electroweak asymmetry 
in the elastic scattering of polarized electrons from the 
proton.   The result is $A=-15.05\pm0.98(stat)\pm0.56(syst)$ ppm at
the kinematic point $\langle\theta_{\mathrm lab}\rangle= 
12.3^{\circ}$ and  
$\langle Q^2\rangle=0.477 $ (GeV/c)$^2$.
Both errors are a factor of two smaller than those of the result
reported previously. 
The value for the strange form factor extracted from the data is
$(G_E^s+0.392G_M^s)%/(G_M^p/\mu_p) 
%=0.069\pm0.056\pm0.039$ is extracted from the data,
=0.025\pm0.020\pm0.014$,
where the first error is experimental
and the second arises from the uncertainties in 
electromagnetic form factors.
This measurement is the first fixed-target parity
violation experiment that used either a ``strained'' GaAs
photocathode to produce highly polarized electrons or a Compton
polarimeter to continuously monitor the electron beam polarization.

\end{abstract}


\journal{Physics Letters B}

\end{frontmatter}

It is well known that strange quarks and antiquarks are present in the
nucleon.  An important open question is the role that sea (non-valence) quarks
in general and strange quarks in particular~\cite{GI}
play in the fundamental properties of the
nucleon.  For example, do strange quarks contribute to the charge
radius or magnetic moment of the proton?  If so, the strange form factors
$G_E^s$ and $G_M^s$ are relevant.  A number of papers have
suggested that indeed these form factors may be 
large~\cite{GI,RLJ,HAM,MB,WEI,PARKA,PARKB,DONG,HAM99,HON97}.
Others models suggest small 
contributions~\cite{ITO95,KEOPF,MMSO,MA97}.

Strange form factors can be isolated from up and down quark
form factors by measuring the  parity-violating asymmetry 
$A=(\sigma_R-\sigma_L)/(\sigma_R+\sigma_L)$
in the elastic scattering of polarized electrons from protons~\cite{KM,BM}.
The experiments are challenging since $A\approx A_0\tau\approx 10$
parts per million (ppm).
Here $A_0=(G_FM_p^2)/(\sqrt{2}\pi\alpha)
=316.7$ ppm, where $G_F$ is the Fermi constant for muon decay and
$M_p$ is the
proton mass.  Also $\tau=Q^2/4M^2_p$ where $Q^2$ is the square of the
four-momentum transfer.  Nevertheless,
several experiments have recently published 
%significantly non-zero values 
results for $A$~\cite {SAM,HAPPEX,SAM2}.  
In this letter, we present the most 
precise measurement to date for $A$ of the proton
and determine new limits for the
possible contribution of strange form factors.

Measurements of elastic electromagnetic and electroweak nucleon 
scattering provide three sets of vector form factors.
From this information, the form factors for each flavor
may be determined~\cite{REV}: $G^u_{E,M}$, $G^d_{E,M}$, and $G^s_{E,M}$. 
A convenient alternate set, which is directly accessible
in experimental measurements, is the electromagnetic
form factors $G^{p\gamma}_{E,M},$ $G^{n\gamma}_{E,M}$,
plus $G^0_{E,M}$.  Here $G^0=(G^u+G^d+G^s)/3,$
%The electromagnetic form factors are related to the flavor
%form factors by 
$G^{p\gamma}=\frac{2}{3}G^{u}-\frac{1}{3}G^{d}-\frac{1}{3}G^s,$ 
and $G^{n\gamma}=\frac{2}{3}G^{d}-\frac{1}{3}G^{u}-\frac{1}{3}G^s,$ 
where the last expression assumes charge symmetry.
$G^0$ cannot be accessed in electromagnetic scattering and thus
represents new information on nucleon dynamics that can be accessed only
via measurements of the weak neutral current amplitude.

%Elastic scattering from the nucleon depends on three
%sets of form factors~\cite{REV}, such 
%as $G^{p\gamma}_{E,M},$ $G^{n\gamma}_{E,M}$,
%and $G^0_{E,M}$.  The first two are the electromagnetic form
%factors for the proton and neutron, respectively, and the
%third is a singlet form factor.  Alternatively, the form
%factors may be decomposed into quark flavors: $G^u_{E,M}$,
%$G^d_{E,M}$, and $G^s_{E,M}$.  
%Then $G^0=(G^u+G^d+G^s)/3,$ and
%$G^{p,n\gamma}=\frac{2}{3}G^{u,d}-\frac{1}{3}G^{d,u}-\frac{1}{3}G^s.$ 
%The expression for the neutron assumes charge symmetry.

The theoretical asymmetry in the Standard Model has
a convenient form in terms of $G^0$:  

\begin{equation} 
A_{th}=-A_0\tau\rho'_{eq}\Biggl( 2-4\hat\kappa'_{eq}\sin^2\theta_W-
\frac{\varepsilon\eta_p}{\varepsilon\eta^2_p+\tau\mu_p^2}
\frac{G^0_E+\beta G_M^0}
{(G_M^{\gamma p}/\mu_p)}
\Biggr)-A_A
\label{eq_atheory}
\end{equation}

where $\mu_p(\mu_n)\approx 2.79(-1.91)$ is the proton(neutron) magnetic
moment in nuclear magnetons, 
$\eta_p=\eta_p(Q^2)=G^{p\gamma}_E(Q^2)/(G^{p\gamma}_M(Q^2)/\mu_p),$
$\varepsilon=(1+2(1+\tau)\tan^2\theta/2)^{-1}$ is the 
longitudinal photon polarization,
and $\beta = \tau\mu_p/(\varepsilon\eta_p)$.  
The scattering angle of the electron
in the laboratory is $\theta$.  The contribution
from the proton axial form factor, $A_A=(0.56\pm0.23)$ ppm, is calculated
to be small for our 
kinematics~\cite{MHOL,ZHU00}.  The recent datum from the SAMPLE 
collaboration~\cite{SAM00} is 1.5 standard deviations larger than the 
prediction.~\cite{MHOL,ZHU00}

The parameters $\rho'_{eq}=0.9879$ and $\hat\kappa'_{eq}=1.0029$
include the effect of electroweak radiative corrections~\cite{PDG},
and $\sin^2\theta_W=0.2314$.
If, in addition to $G^0_{E,M}$, the proton and neutron electromagnetic form
factors $G^{p\gamma}_{E,M}$ and $G^{n\gamma}_{E,M}$ are known,
the strange form factors may be determined from

\begin{equation}
G^s_{E,M}=G^0_{E,M}-G^{p\gamma}_{E,M}-G^{n\gamma}_{E,M}.
\label{eq_gs}
\end{equation}

%It is convenient to normalize the form factors to $G^{p\gamma}_M/\mu_p$
%since the normalized form factors depend less on experimental
%uncertainties and tend to vary less with $\tau$.  
%The quantities extracted are
%$G^s_E/(G^{p\gamma}_M/\mu_p)\rightarrow\tau\rho_s$ and 
%$G_M^s/(G^{p\gamma}_M/\mu_p)\rightarrow\mu_s$ for
%the limit $\tau\rightarrow 0$.  Models~\cite{RLJ,HAM,WEI,DONG} suggest
%that the radius parameter $\rho_s$ could be of the order of $\pm2$ and 
%the strangeness contribution to the magnetic moment $\mu_s$ could be
%of the order of -0.3.  If the strange form factors are indeed
%of this scale, our experiment along with other
%experiments in progress should be able to establish their
%presence.

This experiment took place in Hall~A at the Thomas Jefferson 
National Accelerator Facility.  An approximately $35\mu$A beam of 
67-76\% polarized electrons with an energy of 3.3 GeV
scattered from a 15 cm liquid
hydrogen target.  Elastic events were detected by integrating
the signal in 
total-absorption counters located at the focal plane of a pair of
high-resolution magnetic spectrometers.~\cite{HAPPEX,THE00}

\begin{figure}[p] 
\centering
\vspace{-1cm} 
\epsffile{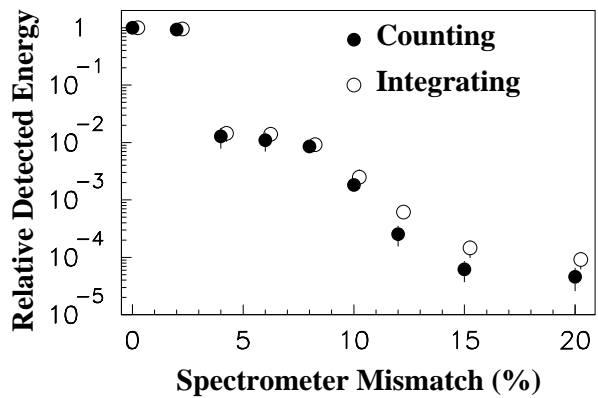} 
%\centerline{\psfig{file=allpairs.eps,width=6in}} 
\caption[pairs]{Fraction of energy deposited in the detector as a function
of spectrometer mismatch.  The inelastic threshold corresponds to a
mismatch of about 4.5\%,
where the response of the detector is already reduced by a factor of 100.}
\label{fig:allp} 
\end{figure}

It is important that the signal be purely elastic,
since background processes may have large asymmetries.
For example, the production of the prominent $\Delta-$resonance 
is calculated to have  3 times the asymmetry of elastic scattering.~\cite{REV}
To measure the rejection of unwanted events by our system,
we measured the response of the detector, both 
in counting and integrating mode, as a function of the mismatch
between the spectrometer setting and the momentum of elastic events.
The result, shown in
Fig. \ref{fig:allp}, is that the integrated response drops many orders of
magnitude as the
momentum mismatch increases.  Based on these data, we determined
that only 0.2\% of our signal arises from inelastic background
processes.  Quasi-elastic scattering from the Al target windows 
contributed 1.5\% to the measured signal.  The net effect of all the
backgrounds is listed in Table \ref{tab:correct}.

\begin{figure}[p] 
\centering
\vspace{-1cm} 
\epsffile{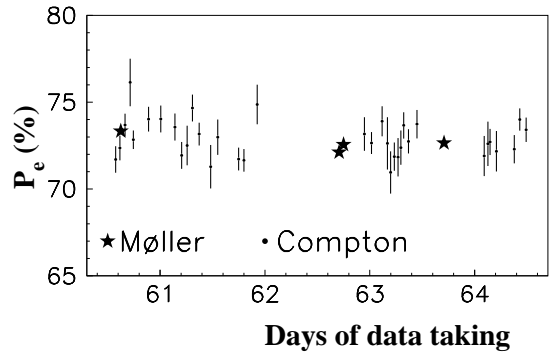} 
\caption[pairs]{Electron beam polarization for part of the run.
The statistical errors on the M{\o}ller data are smaller than the points.}
\label{fig:pol} 
\end{figure}

A new feature of the experiment is that the beam polarization 
$P_e\approx$ 70\%.  This was achieved by using photoemission
by circularly polarized
laser light impinging on a ``strained'' GaAs crystal.  A plot
of the polarization versus time for part of the run
is given in Fig. \ref{fig:pol}.
The starred points are from M{\o}ller scattering and the dots
are preliminary data from the recently commissioned Compton polarimeter.
The errors in the M{\o}ller data have been reduced by a factor of
two from those of Ref. \cite{HAPPEX} by improving our knowledge of the
polarization of the electrons in the magnetized foil target and
our understanding of rate effects in the M{\o}ller spectrometer.
The Compton device continuously monitored the polarization of the beam on 
target and ruled out possible significant variations in polarizations
between the daily M{\o}ller measurements.  
%This data represents the first
%results from a Compton polarimeter in an external beam.
Both devices have an overall systematic error $\Delta P_e/P_e\sim3.2\%.$

\begin{figure}[p] 
\centering

\vspace{-1cm} 
\epsffile{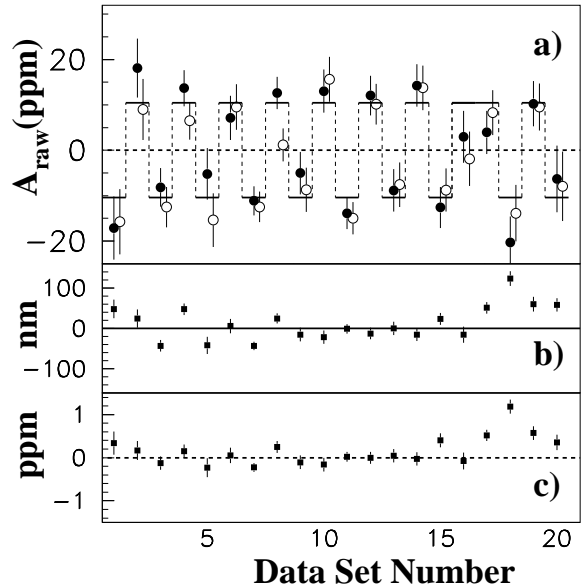} 
\caption[pairs]{a) Raw asymmetry versus data set.  Solid(open) circles
are from the left(right) spectrometer.
The step pattern is due to the insertion of the half-wave plate.
The $\chi^2= 33.7$ for
39 degrees of freedom.  b) Helicity-correlated horizontal position
difference measured near the target.  c) Correction to left
spectrometer data due to all of the beam parameter differences.  The
corrections for the right spectrometer are smaller.}
\label{fig:rawa} 
\end{figure}

To study possible systematic errors in our small asymmetry, we
sometimes inserted a second half-wave $(\lambda/2)$ plate in the 
laser beam at the source
to reverse the sign of the helicity.  Data were obtained in sets
of 24-48 hour duration, and the state of the $\lambda/2$ plate
was reversed for each set.  The resulting asymmetries are shown
in Fig. \ref{fig:rawa}a.  The asymmetry reverses as expected but otherwise 
behaves statistically.

The strained GaAs crystal, in contrast to the bulk GaAs used
for our previous work~\cite{HAPPEX}, has a large analyzing power for linearly
polarized light.~\cite{MAI96}  The consequence was a tendency for 
much larger helicity-correlated differences
in the beam position.  We found that an additional half-wave
plate in the laser beam reduced this problem to a manageable level.
In addition, the intensity asymmetry
of the beam in another experiental hall was nulled to prevent beam
loading in the accelerator from inducing position correlations in our beam.
The remaining position and energy differences 
were measured  with precision microwave monitors.
One example of monitor data is shown in Fig. \ref{fig:rawa}b.  
The effect of these
beam differences on the asymmetry was measured by calibrating the 
apparatus with beam correction coils and an energy 
vernier.  The resultant correction, shown in 
Fig. \ref{fig:rawa}c, proved to have an average of $0.02\pm0.02$ ppm.

The experimental asymmetry, corrected for the measured beam polarization,
is $A_{exp}=-15.1$ at $Q^2=0.477$ (GeV/c)$^2$ for the 1999 data.
We also include the previously reported 1998 data,~\cite{HAPPEX} 
which gives $A_{exp}= -14.7$ ppm 
when extrapolated to the same $Q^2$ value but with approximately
twice the statistical and systematic errors.  
In addition, three small corrections based on subsequent data analysis
were made to the 1998 data:  i) the background correction was included;
ii) the measured beam polarization was reduced by 1.5\%; 
and iii) the $Q^2$ value was determined to be 0.474 (GeV/c)$^2$
instead of 0.479 (GeV/c)$^2$.
An increase of 1\% in $Q^2$ is expected to increase the magnitude of 
the asymmetry  by 1.5\%.  
The errors for the full data set are given in Table \ref{tab:correct}.
Systematic errors in the beam polarimetry and in the measurement
of the spectrometer angle were the most significant sources.
The combined result is $A_{exp}-15.05\pm0.98(stat)\pm0.56(syst)$ ppm at
the average kinematics $Q^2$=0.477 (GeV/c)$^2$ and
$\theta=12.3^{\circ}$.
%To account for the averaging over the finite solid angle of the spectrometers,
%increasing the measured asymmetry was increased by 0.7\% over the value 
%at the nominal kinematics.
This is the average asymmetry over the finite solid angle of the spectrometers;
we estimate the value at the center of acceptance is smaller by 0.7\%.
%We use $A_A=(0.56\pm0.23)$ ppm~\cite{PDG,MHOL,ZHU00}, where the
%uncertainty comes from weak radiative corrections.  

By using Eq. \ref{eq_atheory} and the theoretical value 
for $A_A$~\cite{MHOL,ZHU00}, we obtain 
$(G^0_E+\beta G^0_M)/(G_M^{p\gamma}/\mu_p)
=1.527\pm0.048\pm0.027\pm0.011$. Here the
first error is statistical, the second systematic, and the last error
is due to the uncertainty from $A_A$.  For our kinematics $\beta = 0.392$.
The sensitivity to $\eta_p$ is negligible.  To determine
the contribution due to strange form factors, we use Eq. \ref{eq_gs}
and data for the electromagnetic form factors.  The values we
use~\cite{WAL94,JON00,ANK98,PAS99,HER99,OST99,BEC99,ROH99} are
summarized in Table \ref{tab:emff}.  Thus we
have $G^s_E+\beta G^s_M
%/(G_M^{p\gamma}/\mu_p)
%=0.069\pm0.056\pm0.039$, 
=0.025\pm0.020\pm0.014$, 
where the first
error is the errors in $G^0$ combined in quadrature and the second due
to the electromagnetic form factors.  
This value is consistent with the hypothesis that the
strange form factors are negligible.

We note that there are data for $G_M^n$~\cite{BRU95} that are less precise but
at variance with those of Ref.~\cite{ANK98}.  Our
result for $G^s_E+\beta G^s_M$ would increase by 0.020 if the
data from Ref.~\cite{BRU95}
were used.  New data for both $G_M^n$ and $G_E^n$ are
in the early stages of analysis and will be important both for 
validating our choices and also for
interpreting future data on strange form factors.

\begin{figure}[p] 
\centering
\vspace{-1cm} 
\epsffile{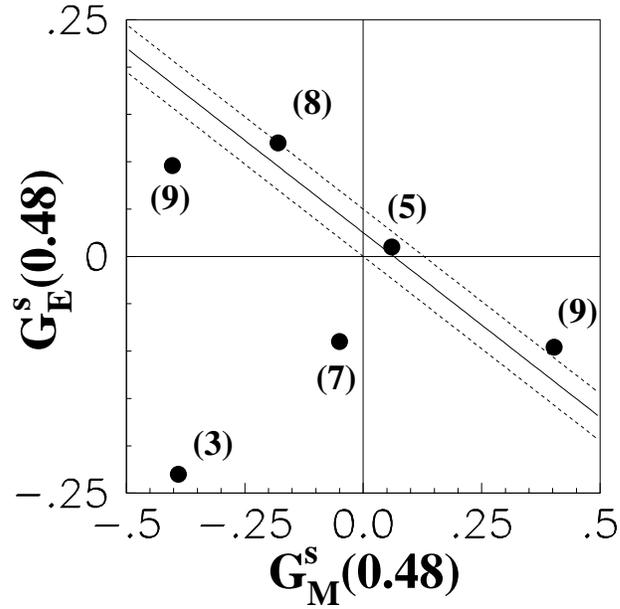} 
%\centerline{\psfig{file=allpairs.eps,width=6in}} 
\caption[pairs]{Plot of $G_E^s$ versus $G_M^s$ at $Q^2=0.477$ (GeV/c)$^2$. 
The band is the allowed region derived from our results.
The width of the band is computed by adding the errors in quadrature.
The points are various estimates from models that make predictions at our
value of $Q^2$.  The numbers in the brackets are the reference of
the models.  Ref. [9] is plotted twice due to an ambiguity
in the predicted sign.}
%\cite{GI,RLJ,HAM,MB,WEI,PARKA,PARKB,DONG,HAM99,HON97,ITO95,KEOPF,MMSO,MA97}
%Open circles are extrapolations from low $Q^2$ of models that predict
%small strange form factors.}
%\cite{GI,RLJ,HAM,MB,WEI,PARKA,PARKB,DONG,HAM99,HON97}}

\label{fig:band} 

\end{figure}

In Fig. \ref{fig:band}, we plot the above value for $G^s_E+\beta G^s_M$
as a band with the errors added in quadrature.
The dots represent the predictions from those models
that apply at our value of $Q^2$.
Our result restricts significantly the possible ``parameter space'' for 
strangeness to be an
important degree of freedom in nucleon form factors.
However, our data are compatible with several models
that predict large strange form factors, including two
with $G_E^s\approx -0.39G_M^s$,~\cite{DONG,HAM99} and one where
the prediction happens to cross zero near our $Q^2$ value.\cite{WEI}

%If we assume that the $\tau\rightarrow0$ limit for the ratio of form factors
%is valid at our $Q^2$, we obtain $\rho_s+2.9\mu_s=0.51\pm0.41\pm0.29$.  
%This result with the quoted errors added in quadrature
%is plotted in Fig. \ref{fig:band}, together with various
%predictions.  The allowed region in parameter space is severely
%restricted, and some predictions are clearly excluded.~\cite{RLJ,HAM}
%However, several models with substantial strange form factors
%are compatible with our result.~\cite{PARKA,DONG,HAM99}%

Our collaboration has two new experiments
approved at JLab for a kinematic point at $Q^2\sim0.1$ (GeV/c)$^2$.
One, using a hydrogen target, will measure the same combination of
strange form factors at a low $Q^2$~\cite{KUM99} and the other, 
using a $^4$He target,
will be sensitive to $G_E^s$ but not $G_M^s$.~\cite{ARM00}
Thus these experiments might detect the presence
of strange form factors that cannot be excluded by the
present result.

%A different assumption for the $Q^2$-dependence, suggested
%by the Galster approximation~\cite{REV} to $G_E^{n\gamma}$, is that
%$G^s_E/(G_M^{p\gamma}/\mu_p)=\tau\rho_s/(1+\lambda^s_E\tau)$,
%where $\lambda_E^s\approx 5.6$.  This would reduce our sensitivity 
%to $\rho_s$ by about a factor of two.  On the other hand, if $\lambda_E^s$
%is negative, our sensitivity is increased.  
%The electromagnetic form factors are a major source of
%uncertainty for $(G^s_E+\beta G^s_M)$.  Moreover, there exist
%for $G_M^n$ data~\cite{BRU95}, also given in Table \ref{tab:emff}, that are
%inconsistent with the value we chose.  With this choice,
%$(G^s_E+\beta G^s_M)/(G_M^{p\gamma}/\mu_p)=0.122\pm0.056\pm0.047$.
%Fortunately, there are experiments in progress that will
%significantly improve the accuracy of the electromagnetic form factors.
%These new electromagnetic measurements could have significant
%impact on the conclusions that we can draw about strange form factors.
%and to encourage further improvements in measurements of 
%the weak form factors.

We wish to thank the entire staff at JLab for their tireless 
work in developing this new facility, and particularly 
C. K. Sinclair and M. Poelker for their timely work on the polarized 
source.  This work was supported by DOE contract DE-AC05-84ER40150
under which the Southeastern Universities Research Association
(SURA) operates the Thomas Jefferson National Accelerator Facility
and by the Department of 
Energy, the National Science Foundation, the Korean Science 
and Engineering Foundation (Korea), the INFN (Italy), the Natural
Sciences and Engineering Research Council of Canada, the 
Commissariat \`a l'\'Energie Atomique (France), and the Centre National
de Recherche Scientifique (France).

\def\Journal#1#2#3#4{{#1} {#2} (#4) #3}

\def\NCA{{\em Nuovo Cimento} A}

\def\PHYS{{ Physica}}

\def\NPA{{ Nucl. Phys.} A}

\def\MATH{{ J. Math. Phys.}}

\def\PRO{{Prog. Theor. Phys.}}

\def\NPB{{ Nucl. Phys.} B}

\def\PLA{{ Phys. Lett.} A}

\def\PLB{{ Phys. Lett.} B}

\def\PLD{{\em Phys. Lett.} D}

\def\PL{{ Phys. Lett.}}

\def\PRL{Phys. Rev. Lett.}

\def\PREV{\em Phys. Rev.}

\def\PREP{ Phys. Rep.}

\def\PRA{{ Phys. Rev.} A}

\def\PRD{{ Phys. Rev.} D}

\def\PRC{{ Phys. Rev.} C}

\def\PRB{{\em Phys. Rev.} B}

\def\ZPC{{ Z. Phys.} C}

\def\ZPA{{ Z. Phys.} A}

\def\ANNP{ Ann. Phys. (N.Y.)}

\def\RMP{{ Rev. Mod. Phys.}}

\def\CHEM{{\em J. Chem. Phys.}}

\def\INT{{ Int. J. Mod. Phys.} E}

\def\NIM{{ Nucl. Instrun. Methods A}}

\begin{table}[p]
\centering
\caption{Summary of corrections and contributions to the errors in \% 
for the measured asymmetry.}
\begin{tabular}{lccc} 

\hline
Source          &Correction (\%)        & $\delta A/A$(\%):1998 
&$\delta A/A$(\%):1999\\
Statistics      &$-$    & 13.3  &7.2\\
$P_{e}$ &$-$    & 7.0   &3.2\\
$Q^2$           &$-$    &1.8    &1.8\\
Backgrounds     &1.2    & 0.6   &0.6\\
\hline
\end{tabular} 
\label{tab:correct}
\end{table}

\begin{table}[p]

\centering
\caption{Electromagnetic form factors normalized to $G_M^p/\mu_p$.
The last column is the error in
$A_{th}$ from the quoted error in the corresponding form factor.}
\label{tab:emff}
\begin{tabular}{lccc} 
\hline
Form Factor     &Value  &Ref. &$\delta A_{th}/A_{th}$\\
 $G_E^p/ (G_M^p/\mu_p$) & $0.99\pm0.02$&\cite{WAL94,JON00}&3\%\\
 $G_E^n/(G_M^p/\mu_p$)  & $0.16\pm0.03$
&\cite{PAS99,HER99,OST99,BEC99,ROH99}&4\% \\
 $(G_M^n/\mu_n)/(G_M^p/\mu_p$)  & $1.05\pm0.02$&\cite{ANK98}&2\%\\  \hline
%&&&\\ 
% $(G_M^n/\mu_n)/(G_M^p/\mu_p$) & $1.12\pm0.04$&\cite{BRU95}&4\%\\
%\hline

\end{tabular}

\end{table}

\epsfxsize=9cm 

\epsfxsize=8cm 

\epsfxsize=9cm 

\newpage

\epsfxsize=9cm 

\end{document}